# Doping-free Janus homojunction solar cell with efficiency exceeding 23%

Lei Li[1], Zi-Xuan Yang[1], Tao Huang[1], Hui Wan[2,1], Wu-Yu Chen[1], Tao Zhang[1], Gui-Fang Huang[1*], Wangyu Hu[3], Wei-Qing Huang[1*]

[1]*Department of Applied Physics, School of Physics and Electronics, Hunan University, Changsha 410082, China*
[2]*School of Materials and Environmental Engineering, Changsha University, Changsha 410082, China*
[3]*School of Materials Science and Engineering, Hunan University, Changsha 410082, China*

**Abstract**: Photovoltaic solar cell is one of the main renewable energy sources, and its power conversion efficiency (PCE) is improved by employing doping or heterojunction to reduce the photogenerated carrier recombination. Here, we propose a doping-free homojunction solar cell utilizing two-dimensional Janus semiconductors to achieve high PCE. Thanks to the intrinsic dipole of Janus structure, doping-free Janus homojunction has naturally not only a type-II band alignment to promote the photoexciton dissociation, but also a smaller effective bandgap to enhance light absorption. More importantly, the intrinsic electric field across the Janus structure will drive photoinduced electron and hole transfer from the interface to the opposite transport layers respectively, significantly enhancing the efficiency of carrier separation and transport. We illustrate the concept in titanium-based Janus monolayer homojunction, where the theoretically observed PCE reaches 23.22% of TiSSe homojunction. Our work opens a novel avenue to design low-cost, high-efficiency solar cells.

---
*.Corresponding authors: wqhuang@hnu.edu.cn, gfhuang@hnu.edu.cn

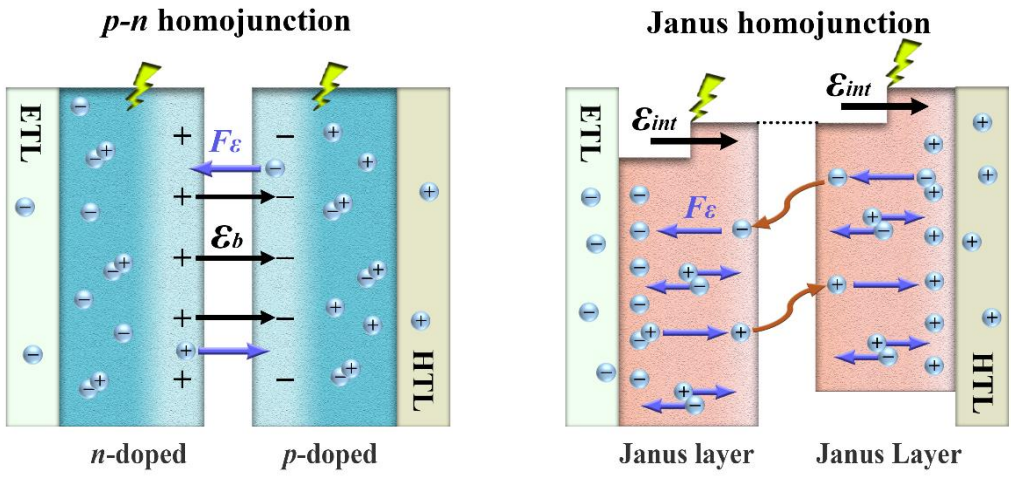

**Table of Contents (TOC)**

Solar energy is the largest carbon-neutral energy source available today and the only fully renewable source that has the capability to meet the world's large and growing energy demand.[1,2] Photovoltaic solar cell, one of the most promising and efficient technologies to harvest this energy, can directly convert sunlight into electricity with zero operating carbon emissions.[3,4] The core of a typical solar cell consists of a photoactive layer (as light absorber) sandwiched by an electron-transport layer (ETL) and hole-transport layer (HTL) (Figure 1).[5] The photoactive layer functions to convert photons into free charge carriers, while the ETL and HTL are crucial for separating and extracting these photogenerated carriers. Moreover, they block counter charge carriers and transport the charge carriers to the electrodes, resulting in photovoltage and photocurrent. Power conversion efficiency (PCE) is key for a solar cell,[6] which will determine whether it is competitive.

Both the materials and configurations of the photoactive layer lie at the heart of improving PCE. Various materials, including silicon,[5] compound semiconductors,[7,8] polymers,[9,10] organic dyes,[11] quantum dots,[12] and perovskites,[13,14] have been identified as suitable photoactive components for solar cells. Concurrently, to effectively separate the photo-generated carriers, several innovative configurations of the photoactive layer have been proposed. Single-junction configuration is used in a bulk heterojunction solar cell, as shown in Figure 1a.[10,15] However, the PCE of single-junction solar cell is still limited by non-ideal exciton dissociation and charge transport, because it is difficult to control the morphology of the phase separation into an interpenetrating network during the thin-film deposition process. Type-II heterojunction configuration can effectively separate the photo-generated carriers to different layers (Figure 1b).[16-19] However, the conversion efficiency in such a bilayer heterojunction device is still limited: (1) Efficient charge separation occurs only at the interface, while the photo-excitations created far from the interface will recombine before diffusing to the interface; (2) Even if charges are separated at the interface, the conversion efficiency is limited by the possible recombination of carriers during long distance transport before reaching transport layers. To reduce the carrier losses, a p-n homojunction structure is proposed (Figure 1c), in which a doping-induced built-in electric field is formed that could enhance the oriented transport of the photo-induced electrons/holes.[20-22] To a certain degree, however, the doping homojunction structure has similar issues as in single junction and type-II heterojunction configurations. Even worse, the doped impurities will act as carrier recombination centers, thus increasing carrier losses. It naturally raises the question of whether there is a strategy to address these drawbacks.

Here, we demonstrate a doping-free Janus homojunction structure consisting of two-dimensional (2D) Janus semiconductors, as a photoactive component for solar cells, to achieve high PCE (Figure 1d). The intrinsic dipole induced by inversion symmetry-breaking[23] naturally results in a type-II band alignment in the Janus homojunction, facilitating the separation of photoexcitons at the interface. Additionally, the interfacial interaction leads to a reduced effective bandgap, enhancing light absorption. In particular, the intrinsic electric field in the Janus layer would drive the oriented transport of the photoinduced electrons and holes in opposite directions, thereby significantly reducing the carrier recombination losses. The results show that the Janus homojunction is an effective photoactive component, beyond existing heterojunction and doping homojunction, for solar cell to achieve high PCE with reduced carrier recombination losses by intrinsic properties.

All the calculations are performed using the first-principles theory within the framework of DFT as implemented in the Vienna Ab Initio Simulation Package (VASP).[24,25] The projector augmented wave (PAW) potentials are utilized to deal with the interaction between core electrons,[26,27] and the plane wave cutoff energy is set to be 500 eV. Generalized gradient approximation (GGA) given by Perdew-Burke-Ernzerhof (PBE) is adopted for exchange-correlation functionals.[28] The atomic coordinates are fully relaxed to ensure the total energy in $10^{-5}$ eV precision and force in $10^{-2}$ eV Å$^{-1}$ convergence, and for the self-consistent field (SCF) calculation the total energy accuracy is increased to $10^{-7}$ eV. The $k$-points with a $9 \times 9 \times 1$ grid for structure relaxation and a $12 \times 12 \times 1$ grid for the SCF calculations are sampled by the Gamma-centered Monkhorst–Pack $k$-grid scheme.[29] The screened hybrid Heyd–Scuseria–Ernzerhof (HSE06) functional is used for the calculations of energy band with more reliable bandgaps.[30,31] The dipole correction is considered in the dipole moments calculation. The long-range interaction is described by the Grimme's DFT-D3 model with Becke-Johnson damping.[32-34] A vacuum space of 30 Å is inserted along the z direction to eliminate the coupling force between the periodic images. The upper limit of PCE for an excitonic solar cell is estimated using the approach developed by Scharber et al.[6] as follows:

$$\eta = \frac{J_{SC}V_{OC}\beta_{FF}}{P_{solar}} = \frac{0.65(E_g^d - E_c - 0.3)\int_{E_g}^{\infty}\frac{P(\hbar\omega)}{\hbar\omega}d(\hbar\omega)}{\int_0^{\infty}P(\hbar\omega)\hbar\omega} \quad (I)$$

where the maximum open circuit voltage $V_{OC}$ is expressed by $(E_g^d - E_c - 0.3)$, and $E_g^d$ and $E_c$ represent the bandgap of donor and conduction band offset of the heterostructure, respectively; the band-fill factor $\beta_{FF}$ is the empirical value derived by Shockley–Queisser, which is assumed to be 0.65; $P(\hbar\omega)$ represents the AM 1.5 solar energy flux at the photon energy, which is defined as $\hbar\omega$. The integration in the numerator and denominator represent the short circuit current and the sum of AM1.5 incident solar irradiation, respectively. $E_g^d$ is regarded as the lower limit of integration of term $\int_{E_g}^{\infty}\frac{P(\hbar\omega)}{\hbar\omega}d(\hbar\omega)$, where the electrons would be excited into higher level after absorbing photon energy over $E_g^d$.

To effectively separate photogenerated carriers, a type-II band alignment is generally adopted for exciton solar cells, as illustrated in Figure 2a. In such an alignment, the PCE can be derived as a function of the bandgap of the donor layer ($E_g^d$) and the conduction-band offset of the heterostructure ($E_c$), as shown in Figure 2b. Ideally, a solar cell can achieve a PCE exceeding 25% when $E_g^d$ is near 1.5 eV and $E_c$ approaches 0 eV. Customizing the band edges to fulfill these criteria is a complex task for a heterojunction or *p-n* homojunction. Here, we propose that constructing Janus homojunction is an effective strategy to remarkably simple this issue. Within the Janus homojunction depicted in Figure 2c, $E_c$ is equal to electrostatic potential difference ($\Delta V$), because the misalignment comes from the dipole induced $\Delta V$; $E_g^d$ is equal to bandgap ($E_g$) of the monolayer, because two identical layers obviates the need to distinguish between donor and acceptor. Therefore, by utilizing a Janus monolayer with $E_g$ of approximately 1.5 eV and $\Delta V$ approaching 0, we can effectively construct a homojunction that could realize high PCE.

To demonstrate our proposal, 2H-phase transition metal dichalcogenides (TMDs) monolayers are chosen as models due to their robust stability as semiconductor materials.[35-37] Their Janus counterparts *MXY* (where *M* represent center metal atom; *X*, *Y* = O, Se, Se, Te; *X* is more electronegative than *Y*), not only inherit semiconductor properties but also introduce an additional intrinsic dipole.[38,39] The calculated dipole moments are illustrated in Figure 2d, with negative values indicating an abnormal direction. The positive direction of the dipole is defined as pointing from the side with lower electronegative atom (*Y*) to another side (*X*). It is observed that all Ti-based materials, as well as Zr- and Hf-based materials that lack oxygen, exhibit minimal dipole moments, with the majority displaying an anomalous direction. To explain the phenomenon, we simplify the dipole ($\mu$) in three-atom-layer structure as two local dipoles ($\mu_X$ and $\mu_Y$) which are from center metal to each side, as shown in Figure 2e, and the $\mu$ depends on the difference between them.[23] The $\mu_X$ and $\mu_Y$ are related to the charge transfer $q_i$ and the distance $d_i$ of the corresponding side:

$$\mu_i \sim q_i \cdot d_i; (i = X, Y)$$

Because *X* is more electronegative than *Y*, it attracts more charge leading to larger $q_X$ than $q_Y$, but simultaneously the *X-M* bond length is shorter than *Y-M* resulting in $d_X$ to be less than $d_Y$, which creates a competitive relationship between $\mu_X$ and $\mu_Y$. In the case $\mu_Y$ larger than $\mu_X$, the direction of dipole would be anomalous which is from *X* to *Y* atoms.

The intrinsic dipole acts like an effective electric field perpendicular to the monolayer plane,[40] inducing a $\Delta V$ on both sides. As shown in Figure 2f, there's a linear correlation between $\mu$ and $\Delta V$. Here, to minimize the $E_c$, Ti*XY* (*X*, *Y* = S, Se, Te) monolayers which possess small dipoles are selected to construct homojunction, as displayed in Figure 3a, with their lattice constants listed in Table 1. For solar cell efficiency, the appropriate $E_g$ is crucial. The energy band calculations reveal that all three structures are semiconductors, as illustrated in Figure 3b. Specifically, TiSSe possesses a moderate $E_g$ of 1.59 eV, ideal for homojunction solar cells. However, TiSeTe and TiSTe possess $E_g$ of 0.88 and 0.78 eV, respectively, which are too small for solar cells.

For comparative study, three Ti*XY* monolayers are all used to construct homojunction. Here,

five high-symmetry stacking styles denoted by *AA*, *AA′*, *A′B*, *AB′*, and *AB* are considered, as depicted in Figure 3c.[41,42] The layers are defined by the ligands at the interface, termed layer-*X* and layer-*Y*. To find the most preferred stacking pattern, the energetic stability is assessed through the binding energy ($E_b$):

$$E_b = E_{hj} - 2E_{mono}$$

where $E_{hj}$, and $E_{mono}$ are the energies of corresponding homojunction and monolayer. The *A′B* stacking exhibits the highest $E_b$, and the energy differences relative to the *A′B* stacking for other stacking are summarized in Figure 3d. The *AA* and *A′B* stackings, with axially aligned ligands, experience interfacial repulsion, leading to higher $E_b$. In contrast, the *AA′*, *AB*, and *AB′* stackings feature an axially staggered arrangement that fosters tighter structural bonding due to enhanced van der Waals interactions, especially for the *AB′* stacking, with the lowest $E_b$ values of -7.26 eV, -6.10 eV, and -5.52 eV for TiSSe, TiSeTe, and TiSTe, respectively. Consequently, the *AB′* stacking is chosen for the homojunction in this study. Figure 3e shows that there is a significant transfer of charge density from the layer-*Y* to layer-*X* under the *AB′* stacking, indicating strong interlayer charge transfer and stable bonding.

The photogenerated carrier separation is firstly tested. Figure 4a shows the density of states projected to two layers of the homojunctions. As expected, all homojunctions exhibit a type-II alignment, where the layer-*Y* contributing valence band maximum (VBM) acts as the donor, and the layer-*X* contributing conduction band minimum (CBM) acts as the acceptor. The wave functions at VBM and CBM distributed in layer-*Y* and layer-*X* respectively, as shown in Figure 4b, further confirms the effectively carrier separation capability. As a promising light absorber in solar cell, the ability of solar light harvesting is essential as well. Thus, we examine the optical properties by calculating the absorption spectra for each monolayer and homojunction, as illustrated in Figure 4c. The homojunctions demonstrate superior absorption in the visible-light region compared to their constituent monolayers, attributed to the reduced effective bandgap ($E_g^e$) upon homojunction formation.

The approach which achieves a type-II band alignment and strong solar light harvesting capacity, is attributed to band bending induced by *ΔV* from the intrinsic dipole, and the further reduced $E_g^e$. In addition to the intrinsic dipole, there is an interlayer dipole in each homojunction due to charge transfer between two layers (Figure 3e). This interlayer dipole induces an additional interlayer electrostatic potential difference (*ΔV*<sub>int</sub>), leading to a shift in the energy bands, as shown in Figure 5a-c.

Unlike intrinsic dipoles, interlayer dipoles are solely associated with the charge transfer between the two ligand atoms at the interface. The direction is always from layer-*Y* pointing towards layer-*X* due to charge transfer occurring from the *Y* to *X* atoms. The intrinsic dipole direction, however, is not fixed, and only aligns with the interlayer dipole direction under conditions of anomalous intrinsic dipole direction. In this case, $E_c$ equals the sum of the individual electrostatic potential differences induced by both intrinsic and interlayer dipoles (*ΔV*+*ΔV*<sub>int</sub>), as illustrated in the

TiSSe homojunction depicted in Figure 5a. For TiSTe with a normal dipole direction, the intrinsic dipole and the interlayer dipole point in opposite directions. Consequently, $E_c$ is determined by the difference between the electrostatic potential differences produced individually by two dipoles ($|ΔV-ΔV_{int}|$), as shown in Figure 5c. Since $E_c$ in this case is a difference, it would obviously be smaller, which is beneficial to improve the PCE. For TiSeTe, where the intrinsic dipole is nearly zero, $E_c$ is then solely contributed by $ΔV_{int}$ (Figure 5b).

The charge transfer not only induces an additional $ΔV_{int}$ but also causes intralayer charge redistribution, which affects the intrinsic dipole of each layer and modifies the $E_g$. Nonetheless, the former undergoes changes that are far less significant compared to the interlayer dipole, and the latter is influenced by more factors. For donor layer (layer-$X$), interlayer dipole acts like an effective electric field applied to the $X$ atom side, leading to a significant change in its bandgap due to the Stark effect.[43,44] As shown in Figure 5d-f, under the synergistic effects of charge redistribution and the Stark effect, the donor bandgap decreases to 1.36, 0.69, and 0.61 eV for TiSSe, TiSeTe, and TiSTe homojunctions, respectively.

Based on $E_c$ and $E_g^d$ values, we can calculate ratio of short circuit current density to AM1.5 solar energy flux $J_{sc}/P_{solar}$ and PCE $η$ to measure the ability to convert photon energy into electric current, as listed in Table 1. The calculated PCE of TiSeTe and TiSTe homojunction are 9.08% and 8.86%, respectively, which are relatively low for applications. However, the TiSSe homojunction exhibit a significantly higher PCE up to 23.22%, which is much closer to the optimal efficiency target for solar cells. In particular, for TiSSe case, the intrinsic electric field in the Janus layer would drive the oriented transport of the photoinduced electrons and holes to directions of ETL and HTL, respectively (Figure 5a), promoting the carrier transfer to the electrode.

In summary, we propose a strategy to design high PCE solar cells with doping-free Janus homojunction based on 2D Janus semiconductors, exemplified by titanium-based monolayers like TiSSe. Consequently, our homojunctions inherently feature a type-II band alignment that enhances photoexciton dissociation, and a reduced effective bandgap that enhances light absorption, which can be attributed to the intrinsic dipole of the Janus monolayers. Moreover, the intrinsic dipole induced electric field can drive photoinduced electron and hole transfer from the interface to the opposite transport layers respectively, further enhancing the efficiency of carrier separation and transport. We find that a minimal intrinsic dipole and an optimal bandgap are key to achieve high PCE. The TiSSe homojunction, in particular, achieves an impressive PCE of 23.22%, showcasing the potential of this strategy for developing solar cells that are both high-performing and cost-effective. This work opens new horizons for the creation of advanced solar cell materials with enhanced efficiency and simplified manufacturing processes.

**Acknowledgements**

This work was supported by the National Natural Science Foundation of China (Grants No. 52172088) and Guangdong Basic and Applied Basic Research Foundation (No. 2024A1515010421).

## REFERENCES

(1) Guo, N.; Yu, L.; Shi, C.; Yan, H.; Chen, M. A Facile and Effective Design for Dynamic Thermal Management Based on Synchronous Solar and Thermal Radiation Regulation. *Nano Lett.* **2024,** 24 (4), 1447-1453.

(2) Shi, P.; Li, J.; Song, Y.; Xu, N.; Zhu, J. Cogeneration of Clean Water and Valuable Energy/Resources via Interfacial Solar Evaporation. *Nano Lett.* **2024,** 24 (19), 5673-5682.

(3) Jiang, N.; Zhang, H. W.; Liu, Y. F.; Wang, Y. F.; Yin, D.; Feng, J. Transfer-Imprinting-Assisted Growth of 2D/3D Perovskite Heterojunction for Efficient and Stable Flexible Inverted Perovskite Solar Cells. *Nano Lett.* **2023,** 23 (13), 6116-6123.

(4) Stanton, R.; Trivedi, D. J. Charge Carrier Dynamics at the Interface of 2D Metal-Organic Frameworks and Hybrid Perovskites for Solar Energy Harvesting. *Nano Lett.* **2023,** 23 (24), 11932-11939.

(5) Bati, A. S. R.; Zhong, Y. L.; Burn, P. L.; Nazeeruddin, M. K.; Shaw, P. E.; Batmunkh, M. Next-generation applications for integrated perovskite solar cells. *Commun. Mater.* **2023,** 4 (1), 2.

(6) Scharber, M. C.; Mühlbacher, D.; Koppe, M.; Denk, P.; Waldauf, C.; Heeger, A. J.; Brabec, C. J. Design Rules for Donors in Bulk‐Heterojunction Solar Cells—Towards 10 % Energy‐Conversion Efficiency. *Adv. Mater.* **2006,** 18 (6), 789-794.

(7) Gong, Y.; Zhu, Q.; Li, B.; Wang, S.; Duan, B.; Lou, L.; Xiang, C.; Jedlicka, E.; Giridharagopal, R.; Zhou, Y.; Dai, Q.; Yan, W.; Chen, S.; Meng, Q.; Xin, H. Elemental de-mixing-induced epitaxial kesterite/CdS interface enabling 13%-efficiency kesterite solar cells. *Nat. Energy.* **2022,** 7 (10), 966-977.

(8) Luo, Y.; Chen, G.; Chen, S.; Ahmad, N.; Azam, M.; Zheng, Z.; Su, Z.; Cathelinaud, M.; Ma, H.; Chen, Z.; Fan, P.; Zhang, X.; Liang, G. Carrier Transport Enhancement Mechanism in Highly Efficient Antimony Selenide Thin‐Film Solar Cell. *Adv. Funct. Mater.* **2023,** 33 (14), 2213941.

(9) Yu, H.; Wang, Y.; Zou, X. H.; Yin, J. L.; Shi, X. Y.; Li, Y. H.; Zhao, H.; Wang, L. Y.; Ng, H. M.; Zou, B. S.; Lu, X. H.; Wong, K. S.; Ma, W.; Zhu, Z. L.; Yan, H.; Chen, S. S. Improved photovoltaic performance and robustness of all-polymer solar cells enabled by a polyfullerene guest acceptor. *Nat. Commun.* **2023,** 14 (1), 2323.

(10) He, Z. C.; Xiao, B.; Liu, F.; Wu, H. B.; Yang, Y. L.; Xiao, S.; Wang, C.; Russell, T. P.; Cao, Y. Single-junction polymer solar cells with high efficiency and photovoltage. *Nat. Photonics.* **2015,** 9 (3), 174-179.

(11) Alizadeh, A.; Roudgar-Amoli, M.; Bonyad-Shekalgourabi, S. M.; Shariatinia, Z.; Mahmoudi, M.; Saadat, F. Dye sensitized solar cells go beyond using perovskite and spinel inorganic materials: A review. *Renew. Sust. Energ. Rev.* **2022,** 157, 112047.

(12) Liu, L.; Najar, A.; Wang, K.; Du, M. Y.; Liu, S. Z. Perovskite Quantum Dots in Solar Cells. *Adv. Sci.* **2022,** 9 (7), 2104577.

(13) Tong, Y.; Najar, A.; Wang, L.; Liu, L.; Du, M. Y.; Yang, J.; Li, J. X.; Wang, K.; Liu, S. Z. Wide-Bandgap Organic-Inorganic Lead Halide Perovskite Solar Cells. *Adv. Sci.* **2022,** 9 (14), 2105085.

(14) Zhou, Y. Y.; Herz, L. M.; Jen, A. K. Y.; Saliba, M. Advances and challenges in understanding the microscopic structure-property-performance relationship in perovskite solar cells. *Nat. Energy.* **2022,** 7 (9), 794-807.

(15) Xiao, K.; Lin, R. X.; Han, Q. L.; Hou, Y.; Qin, Z. Y.; Nguyen, H. T.; Wen, J.; Wei, M. Y.; Yeddu, V.; Saidaminov, M. I.; Gao, Y.; Luo, X.; Wang, Y. R.; Gao, H.; Zhang, C. F.; Xu, J.; Zhu, J.; Sargent, E. H.; Tan, H. R. All-perovskite tandem solar cells with 24.2% certified efficiency and area over 1 cm$^2$ using surface-anchoring zwitterionic antioxidant. *Nat. Energy.* **2020,** 5 (11), 870-880.

(16) Huang, T.; Yang, Z.-X.; Li, L.; Wan, H.; Zhang, T.; Huang, G.-F.; Hu, W.; Huang, W.-Q. Symmetry-breaking-enhanced power conversion efficiency of 2D van der Waals heterostructures. *Appl. Phys. Lett.* **2024,** 125 (3), 033901.

(17) Liang, K.; Huang, T.; Yang, K.; Si, Y.; Wu, H.-Y.; Lian, J.-C.; Huang, W.-Q.; Hu, W.-Y.; Huang, G.-F. Dipole Engineering of Two-Dimensional van der Waals Heterostructures for Enhanced Power-Conversion Efficiency: The Case of Janus $Ga_2SeTe$/InS. *Phys. Rev. Appl.* **2021,** 16 (5), 054043.

(18) Si, Y.; Wu, H.-Y.; Yang, K.; Lian, J.-C.; Huang, T.; Huang, W.-Q.; Hu, W.-Y.; Huang, G.-F. High-throughput computational design for 2D van der Waals functional heterostructures: Fragility of Anderson's rule and beyond. *Appl. Phys. Lett.* **2021,** 119 (4), 043102.

(19) Garcia, V. G.; Batista, N. N.; Aldave, D. A.; Capaz, R. B.; Palacios, J. J.; Menezes, M. G.; Paz, W. S. Unlocking the Potential of Nanoribbon-Based $Sb_2S_3$/$Sb_2Se_3$ van-der-Waals Heterostructure for Solar-Energy-Conversion and Optoelectronics Applications. *ACS Appl. Mater. Interfaces.* **2023,** 15 (47), 54786-54796.

(20) Calado, P.; Barnes, P. R. F. Ionic screening in perovskite p–n homojunctions. *Nat. Energy.* **2021,** 6 (6), 589-591.

(21) Cui, P.; Wei, D.; Ji, J.; Huang, H.; Jia, E.; Dou, S.; Wang, T.; Wang, W.; Li, M. Planar p–n homojunction perovskite solar cells with efficiency exceeding 21.3%. *Nat. Energy.* **2019,** 4 (2), 150-159.

(22) Zhao, Z.; Sun, M.; Ji, Y.; Mao, K.; Huang, Z.; Yuan, C.; Yang, Y.; Ding, H.; Yang, Y.; Li, Y.; Chen, W.; Zhu, J.; Wei, J.; Xu, J.; Paritmongkol, W.; Abate, A.; Xiao, Z.; He, L.; Hu, Q. Efficient Homojunction Tin Perovskite Solar Cells Enabled by Gradient Germanium Doping. *Nano Lett.* **2024,** 24 (18), 5513-5520.

(23) Li, L.; Huang, T.; Liang, K.; Si, Y.; Lian, J. C.; Huang, W. Q.; Hu, W. Y.; Huang, G. F. Symmetry-Breaking-Induced Multifunctionalities of Two-Dimensional Chromium-Based Materials for Nanoelectronics and Clean Energy Conversion. *Phys. Rev. Appl.* **2022,** 18 (1), 014013.

(24) Kresse, G.; Furthmüller, J. Efficient iterative schemes forab initiototal-energy calculations using a plane-wave basis set. *Phys. Rev. B.* **1996,** 54 (16), 11169-11186.

(25) Kresse, G.; Hafner, J. Ab initiomolecular-dynamics simulation of the liquid-metal–amorphous-semiconductor transition in germanium. *Phys. Rev. B.* **1994,** 49 (20), 14251-14269.

(26) Blöchl, P. E. Projector augmented-wave method. *Phys. Rev. B.* **1994,** 50 (24), 17953-17979.

(27) Kresse, G.; Joubert, D. From ultrasoft pseudopotentials to the projector augmented-wave method. *Phys. Rev. B.* **1999,** 59 (3), 1758-1775.

(28) Perdew, J. P.; Burke, K.; Ernzerhof, M. Generalized Gradient Approximation Made Simple. *Phys. Rev. Lett.* **1996,** 77 (18), 3865-3868.

(29) Monkhorst, H. J.; Pack, J. D. Special points for Brillouin-zone integrations. *Phys. Rev. B.* **1976,** 13 (12), 5188-5192.

(30) Heyd, J.; Scuseria, G. E. Assessment and validation of a screened Coulomb hybrid density functional. *J. Chem. Phys.* **2004,** 120 (16), 7274-7280.

(31) Heyd, J.; Scuseria, G. E.; Ernzerhof, M. Hybrid functionals based on a screened Coulomb potential. *J. Chem. Phys.* **2003,** 118 (18), 8207-8215.

(32) Grimme, S. Accurate description of van der Waals complexes by density functional theory including empirical corrections. *J. Comput. Chem.* **2004,** 25 (12), 1463-1473.

(33) Grimme, S.; Ehrlich, S.; Goerigk, L. Effect of the Damping Function in Dispersion Corrected Density Functional Theory. *J. Comput. Chem.* **2011,** 32 (7), 1456-1465.

(34) Grimme, S.; Antony, J.; Ehrlich, S.; Krieg, H. A consistent and accurate ab initio parametrization of density functional dispersion correction (DFT-D) for the 94 elements H-Pu. *J. Chem. Phys.* **2010,** 132 (15), 154104.

(35) Sui, X.; Hu, T.; Wang, J.; Gu, B.; Duan, W.; Miao, M. Voltage-controllable colossal magnetocrystalline anisotropy in single-layer transition metal dichalcogenides. *Phys. Rev. B.* **2017,** 96 (4), 041410.

(36) Rasmussen, F. A.; Thygesen, K. S. Computational 2D Materials Database: Electronic Structure of Transition-Metal Dichalcogenides and Oxides. *J. Phys. Chem. C.* **2015,** 119 (23), 13169-13183.

(37) Ataca, C.; Şahin, H.; Ciraci, S. Stable, Single-Layer $MX_2$ Transition-Metal Oxides and Dichalcogenides in a Honeycomb-Like Structure. *J. Phys. Chem. C.* **2012,** 116 (16), 8983-8999.

(38) Dong, L.; Lou, J.; Shenoy, V. B. Large In-Plane and Vertical Piezoelectricity in Janus Transition Metal Dichalchogenides. *Acs Nano.* **2017,** 11 (8), 8242-8248.

(39) Zhang, C.; Nie, Y.; Sanvito, S.; Du, A. First-Principles Prediction of a Room-Temperature Ferromagnetic Janus VSSe Monolayer with Piezoelectricity, Ferroelasticity, and Large Valley Polarization. *Nano Lett.* **2019,** 19 (2), 1366-1370.

(40) Strasser, A.; Wang, H.; Qian, X. F. Nonlinear Optical and Photocurrent Responses in Janus MoSSe Monolayer and $MoS_2$-MoSSe van der Waals Heterostructure br. *Nano Lett.* **2022,** 22 (10), 4145-4152.

(41) Constantinescu, G.; Kuc, A.; Heine, T. Stacking in Bulk and Bilayer Hexagonal Boron Nitride. *Phys. Rev. Lett.* **2013,** 111 (3), 036104.

(42) Zhou, W.; Chen, J.; Yang, Z.; Liu, J.; Ouyang, F. Geometry and electronic structure of monolayer, bilayer, and multilayer Janus WSSe. *Phys. Rev. B.* **2019,** 99 (7), 075160.

(43) Ramasubramaniam, A.; Naveh, D. Piezoelectric electrostatic superlattices in monolayer MoS2. *Phys. Rev. Mater.* **2024,** 8 (1), 014002.

(44) Bera, A.; Maiti, A.; Pal, A. J. Emergence of a hidden topological insulator phase in hybrid halide perovskites. *Appl. Phys. Lett.* **2023,** 123 (13), 133103.

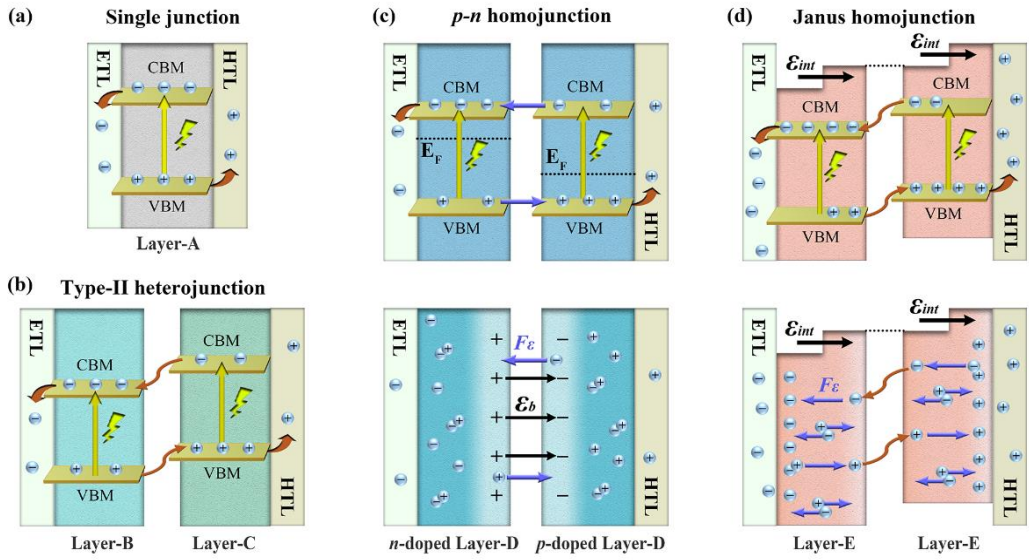

Figure 1. The diagrams of solar cells based on (a) single junction, (b) type-II heterojunction, (c) *p-n* homojunction, and (d) Janus homojunction. The built-in electric field ($\varepsilon_b$) in *p-n* homojunction drives the interlayer carrier migration to different layers, and the intrinsic electric field ($\varepsilon_{int}$) in Janus homojunction drives the carrier migration to opposite directions in each Janus layer. The electric field force ($F_\varepsilon$) on the carrier is marked with blue arrows.

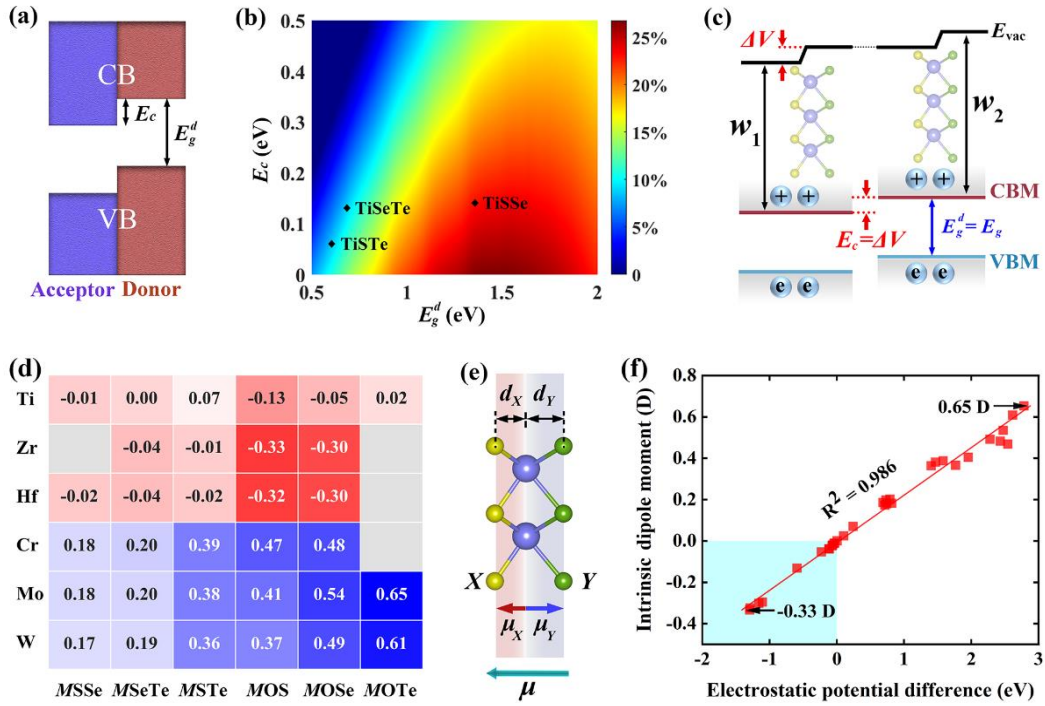

Figure 2. (a) The type-II energy band alignment diagram. (b) Computed PCE contours as a function of the conduction band offset $E_c$ and the donor bandgap $E_g^d$. (c) Band arrangement diagram of Janus homojunction. The (d) dipole moments, (e) dipole diagram, and (f) linear relationship between electrostatic potential differences and intrinsic dipole moments of 2H-phase Janus TMDs monolayers.

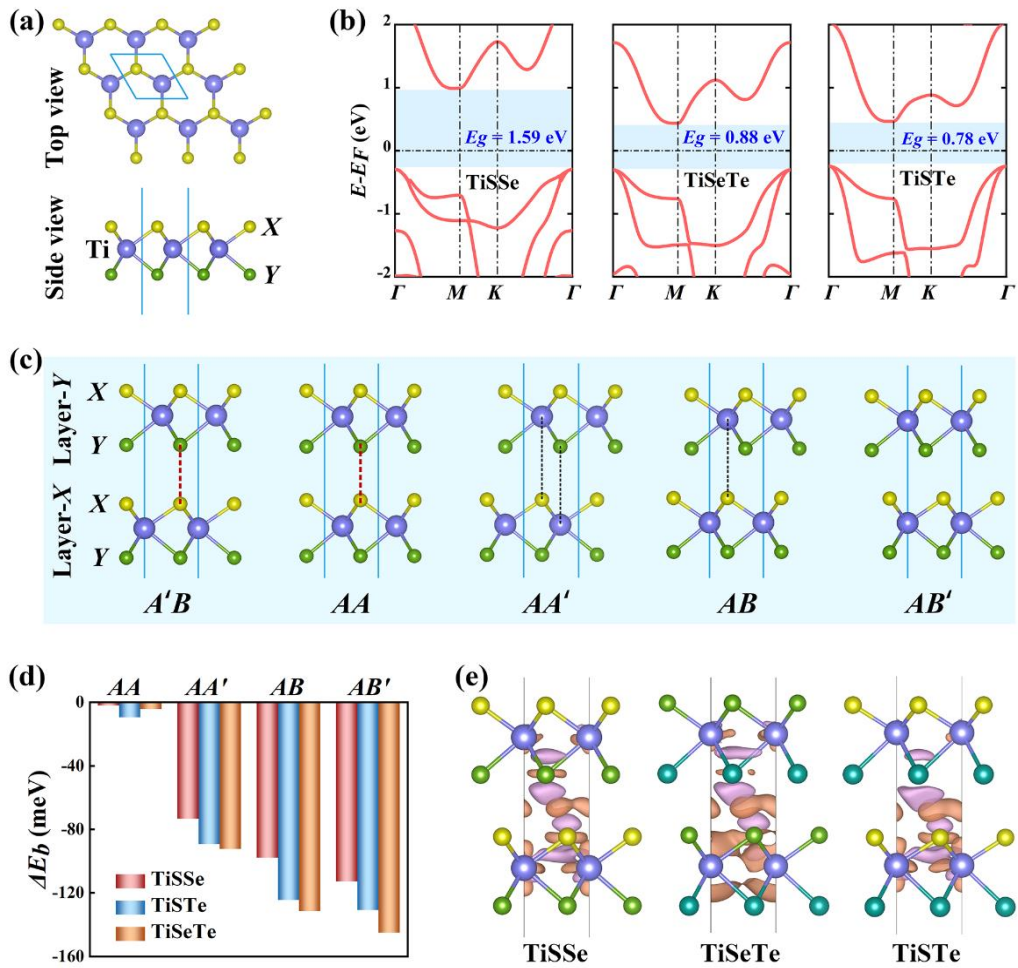

Figure 3. (a) The top and side views of Ti$XY$ ($X$, $Y$ = S, Se, Te) monolayers. (b) The energy bands of Ti$XY$. (c) Diagrams of the stacking styles of Ti$XY$ homojunction. (d) The binding energy differences to $AB'$ stacking. (e) The charge density difference between two layers of Ti$XY$ homojunctions.

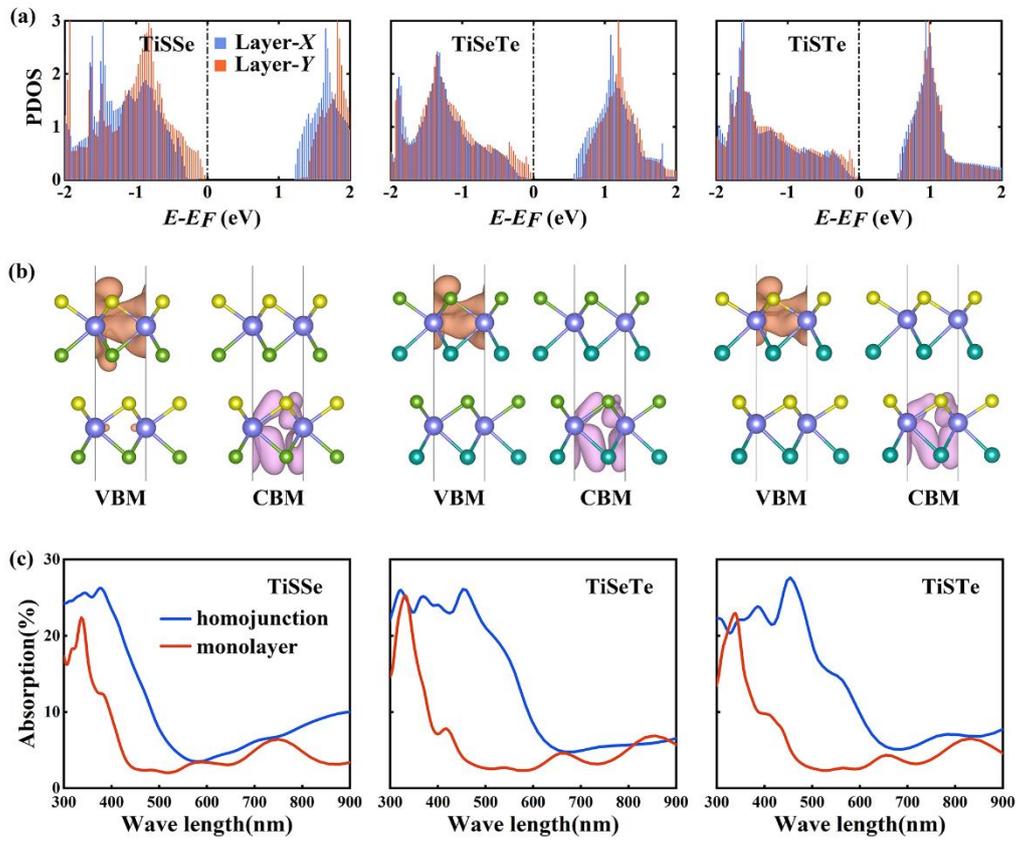

Figure 4. The (a) density of states projected to two layers and (b) corresponding wavefunctions in CBM and VBM of Ti*XY* homojunctions. (c) The absorption coefficients of Ti*XY* monolayers and homojunctions.

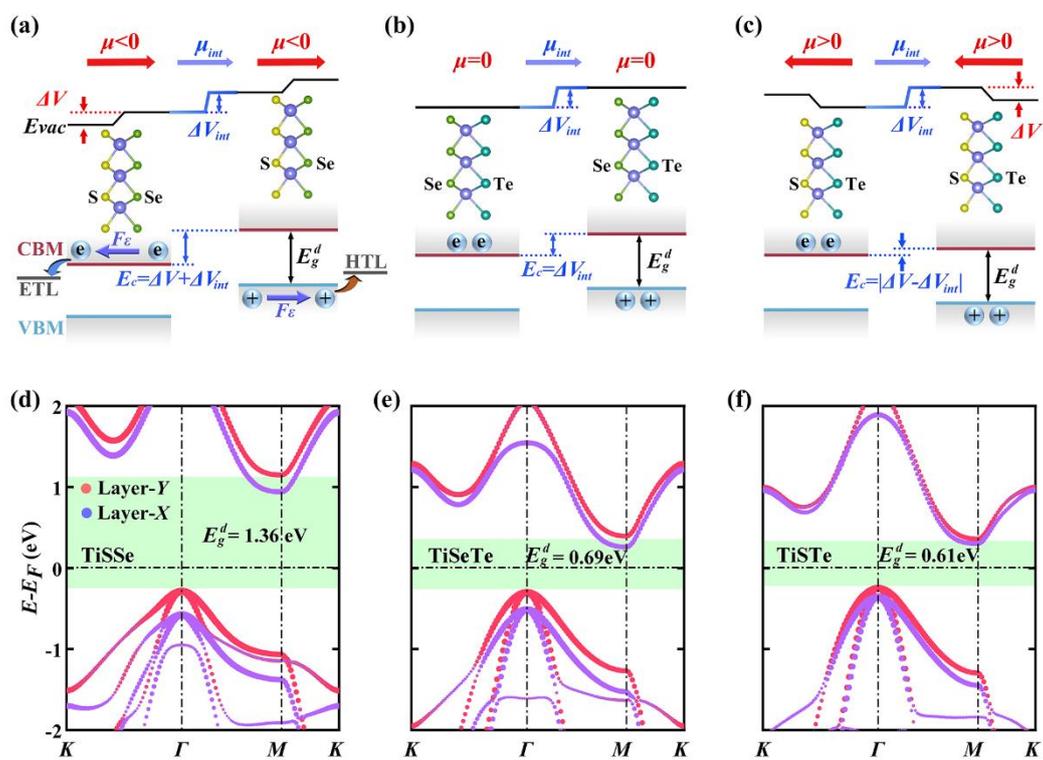

Figure 5. Band arrangement diagram of (a) TiSSe, (b) TiSeTe, and (c) TiSTe homojunctions. The energy bands projected to two layers of (d) TiSSe, (e) TiSeTe, and (f) TiSTe homojunctions.

Table 1. The optimized lattice constants *a*, energy bandgap $E_g$, electrostatic potential difference $\Delta V$ of TiSSe, TiSeTe and TiSTe monolayers; the donor bandgap $E_g^d$, ratio of short circuit current density to AM1.5 solar energy flux $J_{sc}/P_{solar}$, conduction band offset $E_c$, PCE $\eta$ of TiSSe, TiSeTe and TiSTe homojunctions.

|  | monolayer | | | homojunction | | | |
| --- | --- | --- | --- | --- | --- | --- | --- |
|  | $a$(Å) | $E_g$(eV) | $\Delta V$(eV) | $E_g^d$(eV) | $J_{sc}/P_{solar}$ | $E_c$(eV) | $\eta$(%) |
| TiSSe | 3.38 | 1.58 | -0.04 | 1.36 | 0.39 | 0.14 | 23.22 |
| TiSeTe | 3.57 | 0.88 | 0 | 0.69 | 0.54 | 0.13 | 9.08 |
| TiSTe | 3.51 | 0.78 | 0.24 | 0.61 | 0.54 | 0.06 | 8.86 |